\documentclass[12pt]{article}
\usepackage{graphicx}
\usepackage{amsmath}
\usepackage{hhline}
\usepackage{amssymb}
\usepackage{times}
\usepackage{cite}
\usepackage{array}
\usepackage{rotating}
\usepackage{multirow}
\usepackage{bigstrut}
\usepackage{color}
\definecolor{rltred}{rgb}{0.75,0,0}
\definecolor{rltgreen}{rgb}{0,0.5,0}
\definecolor{rltblue}{rgb}{0,0,0.75}

\newif\ifpdf
\ifx\pdfoutput\undefined
    \pdffalse          
\else
    \pdfoutput=1       
    \pdftrue
\fi

\ifpdf
\usepackage{thumbpdf}
\usepackage[pdftex,
        colorlinks=true,
        urlcolor=rltblue,       
        filecolor=rltgreen,     
        linkcolor=rltred,       
        pdftitle={Example File for pdflatex},
        pdfauthor={H1},
        pdfsubject={Template file},
        pdfkeywords={High-Energy Physics, Particle Physics},
        pdfpagemode=None,
        bookmarksopen=true]{hyperref}
 
\fi

\newlength{\dinwidth}
\newlength{\dinmargin}
\begin{document}  

\newcommand{\pom}{{I\!\!P}}
\newcommand{\reg}{{I\!\!R}}
\newcommand{\slowpi}{\pi_{\mathit{slow}}}
\newcommand{\fiidiii}{F_2^{D(3)}}
\newcommand{\fiidiiiarg}{\fiidiii\,(\beta,\,Q^2,\,x)}
\newcommand{\n}{1.19\pm 0.06 (stat.) \pm0.07 (syst.)}
\newcommand{\nz}{1.30\pm 0.08 (stat.)^{+0.08}_{-0.14} (syst.)}
\newcommand{\fiidiiiful}{F_2^{D(4)}\,(\beta,\,Q^2,\,x,\,t)}
\newcommand{\fiipom}{\tilde F_2^D}
\newcommand{\ALPHA}{1.10\pm0.03 (stat.) \pm0.04 (syst.)}
\newcommand{\ALPHAZ}{1.15\pm0.04 (stat.)^{+0.04}_{-0.07} (syst.)}
\newcommand{\fiipomarg}{\fiipom\,(\beta,\,Q^2)}
\newcommand{\pomflux}{f_{\pom / p}}
\newcommand{\nxpom}{1.19\pm 0.06 (stat.) \pm0.07 (syst.)}
\newcommand {\gapprox}
   {\raisebox{-0.7ex}{$\stackrel {\textstyle>}{\sim}$}}
\newcommand {\lapprox}
   {\raisebox{-0.7ex}{$\stackrel {\textstyle<}{\sim}$}}
\def\gsim{\,\lower.25ex\hbox{$\scriptstyle\sim$}\kern-1.30ex%
\raise 0.55ex\hbox{$\scriptstyle >$}\,}
\def\lsim{\,\lower.25ex\hbox{$\scriptstyle\sim$}\kern-1.30ex%
\raise 0.55ex\hbox{$\scriptstyle <$}\,}
\newcommand{\pomfluxarg}{f_{\pom / p}\,(x_\pom)}
\newcommand{\dsf}{\mbox{$F_2^{D(3)}$}}
\newcommand{\dsfva}{\mbox{$F_2^{D(3)}(\beta,Q^2,x_{I\!\!P})$}}
\newcommand{\dsfvb}{\mbox{$F_2^{D(3)}(\beta,Q^2,x)$}}
\newcommand{\dsfpom}{$F_2^{I\!\!P}$}
\newcommand{\gap}{\stackrel{>}{\sim}}
\newcommand{\lap}{\stackrel{<}{\sim}}
\newcommand{\fem}{$F_2^{em}$}
\newcommand{\tsnmp}{$\tilde{\sigma}_{NC}(e^{\mp})$}
\newcommand{\tsnm}{$\tilde{\sigma}_{NC}(e^-)$}
\newcommand{\tsnp}{$\tilde{\sigma}_{NC}(e^+)$}
\newcommand{\st}{$\star$}
\newcommand{\sst}{$\star \star$}
\newcommand{\ssst}{$\star \star \star$}
\newcommand{\sssst}{$\star \star \star \star$}
\newcommand{\tw}{\theta_W}
\newcommand{\sw}{\sin{\theta_W}}
\newcommand{\cw}{\cos{\theta_W}}
\newcommand{\sww}{\sin^2{\theta_W}}
\newcommand{\cww}{\cos^2{\theta_W}}
\newcommand{\trm}{m_{\perp}}
\newcommand{\trp}{p_{\perp}}
\newcommand{\trmm}{m_{\perp}^2}
\newcommand{\trpp}{p_{\perp}^2}
\newcommand{\alp}{\alpha_s}

\newcommand{\alps}{\alpha_s}
\newcommand{\sqrts}{$\sqrt{s}$}
\newcommand{\LO}{$O(\alpha_s^0)$}
\newcommand{\Oa}{$O(\alpha_s)$}
\newcommand{\Oaa}{$O(\alpha_s^2)$}
\newcommand{\PT}{p_{\perp}}
\newcommand{\JPSI}{J/\psi}
\newcommand{\sh}{\hat{s}}
\newcommand{\uh}{\hat{u}}
\newcommand{\MP}{m_{J/\psi}}
\newcommand{\PO}{I\!\!P}
\newcommand{\xbj}{x}
\newcommand{\xpom}{x_{\PO}}
\newcommand{\ttbs}{\char'134}
\newcommand{\xpomlo}{3\times10^{-4}}  
\newcommand{\xpomup}{0.05}  
\newcommand{\dgr}{^\circ}
\newcommand{\pbarnt}{\,\mbox{{\rm pb$^{-1}$}}}
\newcommand{\gev}{\,\mbox{GeV}}
\newcommand{\WBoson}{\mbox{$W$}}
\newcommand{\fbarn}{\,\mbox{{\rm fb}}}
\newcommand{\fbarnt}{\,\mbox{{\rm fb$^{-1}$}}}
%
%
\newcommand{\qsq}{\ensuremath{Q^2} }
\newcommand{\gevsq}{\ensuremath{\mathrm{GeV}^2} }
\newcommand{\et}{\ensuremath{E_t^*} }
\newcommand{\rap}{\ensuremath{\eta^*} }
\newcommand{\gp}{\ensuremath{\gamma^*}p }
\newcommand{\dsiget}{\ensuremath{{\rm d}\sigma_{ep}/{\rm d}E_t^*} }
\newcommand{\dsigrap}{\ensuremath{{\rm d}\sigma_{ep}/{\rm d}\eta^*} }
\def\Journal#1#2#3#4{{#1} {\bf #2} (#3) #4}
\def\NCA{\em Nuovo Cimento}
\def\NIM{\em Nucl. Instrum. Methods}
\def\NIMA{{\em Nucl. Instrum. Methods} {\bf A}}
\def\NPB{{\em Nucl. Phys.}   {\bf B}}
\def\PLB{{\em Phys. Lett.}   {\bf B}}
\def\PRL{\em Phys. Rev. Lett.}
\def\PRD{{\em Phys. Rev.}    {\bf D}}
\def\ZPC{{\em Z. Phys.}      {\bf C}}
\def\EJC{{\em Eur. Phys. J.} {\bf C}}
\def\CPC{\em Comp. Phys. Commun.}

\begin{titlepage}

\begin{flushleft}

DESY 05-110 \hfill ISSN 0418-9833 \\
July 2005
\end{flushleft}

\vspace{2cm}

\begin{center}
\begin{Large}
  
   {\bf {\boldmath {Measurement of $F_2^{c\bar{c}}$ and 
 $F_2^{b\bar{b}}$
      at Low $Q^2$ and $x$ \\ using the H1 Vertex Detector at HERA}}}

\vspace{2cm}

H1 Collaboration

\end{Large}
\end{center}

\vspace{2cm}

\begin{abstract}
  \noindent Measurements are presented of inclusive charm and beauty
  cross sections in $e^+p$ collisions at HERA for values of photon
  virtuality $12 \le Q^2 \le 60~{\rm GeV}^2$ and of the Bjorken
  scaling variable $0.0002 \le x \le 0.005$. The fractions of events
  containing charm and beauty quarks are determined using a method
  based on the impact parameter, in the transverse plane, of tracks to
  the primary vertex, as measured by the H1 vertex detector.  Values
  for the structure functions $F_2^{c\bar{c}}$ and $F_2^{b\bar{b}}$
  are obtained.  This is the first measurement of $F_2^{b\bar{b}}$ in
  this kinematic range. The results are found to be compatible with
  the predictions of perturbative quantum chromodynamics and with
  previous measurements of $F_2^{c\bar{c}}$.

\end{abstract}

\vspace{1.5cm}

\begin{center}
To be submitted to {\em Eur. Phys. J.} {\bf C}
\end{center}

\end{titlepage}

\begin{flushleft}

A.~Aktas$^{10}$,               
V.~Andreev$^{26}$,             
T.~Anthonis$^{4}$,             
S.~Aplin$^{10}$,               
A.~Asmone$^{34}$,              
A.~Astvatsatourov$^{4}$,       
A.~Babaev$^{25}$,              
S.~Backovic$^{31}$,            
J.~B\"ahr$^{39}$,              
A.~Baghdasaryan$^{38}$
% ...

\bigskip{\it
% ...
$ ^{4}$ Inter-University Institute for High Energies ULB-VUB, Brussels;
          Universiteit Antwerpen, Antwerpen; Belgium$^{ c}$ \\
% ...
$ ^{10}$ DESY, Hamburg, Germany \\
% ...
$ ^{25}$ Institute for Theoretical and Experimental Physics,
          Moscow, Russia$^{ l}$ \\
 $ ^{26}$ Lebedev Physical Institute, Moscow, Russia$^{ e}$ \\
% ...
$ ^{31}$ Faculty of Science, University of Montenegro,
          Podgorica, Serbia and Montenegro$^{ e}$ \\
% ...
$ ^{34}$ Dipartimento di Fisica Universit\`a di Roma Tre
          and INFN Roma~3, Roma, Italy \\
% ...
$ ^{38}$ Yerevan Physics Institute, Yerevan, Armenia \\
 $ ^{39}$ DESY, Zeuthen, Germany \\
% ...
$ ^b$ Supported by the UK Particle Physics and Astronomy Research
      Council, and formerly by the UK Science and Engineering Research
      Council \\
 $ ^c$ Supported by FNRS-FWO-Vlaanderen, IISN-IIKW and IWT
      and  by Interuniversity
Attraction Poles Programme,
      Belgian Science Policy \\
% ...
$ ^e$ Supported by the Deutsche Forschungsgemeinschaft \\
% ...
$ ^l$ Partially Supported by Russian Foundation
      for Basic Research, grant    no. 00-15-96584 \\
}

\end{flushleft}

\newpage

\section{Introduction}
Measurements of the charm ($c$) and beauty ($b$) contributions to the
inclusive proton structure function $F_2$ have been made recently in
Deep Inelastic Scattering (DIS) at HERA, using information from the H1
vertex detector, for values of the negative square of the four
momentum of the exchanged boson $Q^2>150$~${\rm
GeV^2}$\cite{Aktas:2004az}. In this high $Q^2$ region a fraction of
$\sim 18 \%$ ($\sim 3 \%$) of DIS events contain $c$ ($b$) quarks. It
was found that perturbative QCD (pQCD) calculations at next-to-leading
order (NLO) gave a good description of the data.  In this paper a
similar method is employed, using data from the same running period,
to extend the measurements to the range of lower $Q^2$, $12
\le Q^2 \le 60$~${\rm GeV}^2$, and of Bjorken $x$, $0.000197 \le x \le 0.005$. 

Events containing heavy quarks are distinguished from those containing
only light quarks by reconstructing the displacement of tracks from
the primary vertex, using precise spatial information from the H1
vertex detector.  The long lifetimes of $c$ and $b$ flavoured hadrons
lead to larger displacements than for light quark events. The charm
structure function $F_2^{c\bar{c}}$ and the beauty structure function
$F_2^{b\bar{b}}$ are obtained from the measured $c$ and $b$ cross
sections after small corrections for the longitudinal structure
functions $F_L^{c\bar{c}}$ and $F_L^{b\bar{b}}$. The measurements at
low $Q^2$ benefit from increased statistics when compared to those at
high $Q^2$. However, the low $Q^2$ region is experimentally more
challenging because the final state does not have as large a
transverse boost in the laboratory frame. The separation between $b$
and $c$ events is also difficult since, although the $c$ fraction is
expected to be similar as at high $Q^2$, the $b$ fraction is expected
to be much smaller ($\sim 0.6\%$ at $Q^2 = 12$~${\rm
GeV}^2$\cite{Martin:2004dh,cteqvfns}).

Previous measurements of the open charm cross section in DIS at
HERA have mainly been of exclusive $D$ or $D^*$ meson
production\cite{H1ZEUSDstar,H1Dstar,ZEUSDstar}.  From the $D^*$ measurements
the contribution of charm to the proton structure function
has been derived by correcting for the fragmentation
fraction $f(c \rightarrow D^*)$ and the unmeasured phase space (mainly
at low values of transverse momentum of the meson). The results are
found to be in good agreement with pQCD predictions.  The $b$ cross
section in DIS, in a similar kinematic region to the present analysis,
has been measured for events containing a muon 
and an associated jet in the Breit frame 
in the final state \cite{zeusBdis, Aktas:2005zc}. 
The measured cross sections are found to be somewhat higher than perturbative 
calculations at NLO.

\section{Theoretical Description of Heavy Flavour Production in Deep Inelastic Scattering}

\subsection{NLO QCD Calculations}
\label{sec:theory}
In the framework of NLO QCD analyses of global inclusive and jet cross
section measurements, the production of heavy flavours is described
using the variable flavour number scheme (VFNS) which aims to provide
reliable pQCD predictions over the whole kinematic range in $Q^2$.  At
values of $Q^2 \simeq m^2$ the effects of the quark mass $m$ must be
taken into account and the heavy flavour partons are treated as
massive quarks.  The dominant LO process in this region is photon
gluon fusion (PGF) and the NLO diagrams are of order
$\alpha_s^2$~\cite{massive}.  As $Q^2$ increases, in the region 
$Q^2 \gg m^2,$ the heavy quark may be treated as a massless
parton in the proton.  Several approaches\cite{VFNS1,VFNS2,VFNS3} have
been developed which deal with the transition from the heavy quark
mass effects at low $Q^2$ to the asymptotic massless parton behaviour
at high $Q^2$. Recently, predictions for inclusive heavy flavour
production within a VFNS approach have been calculated at
next-to-next-to-leading order (NNLO)\cite{NNLO}.

Predictions for the charm and beauty cross sections may also be obtained 
from fits\cite{ccfm} to the HERA inclusive $F_2$ data based on CCFM
evolution\cite{ccfm2}.  The heavy quarks are produced in the fixed flavour
number scheme (FFNS) according to the LO PGF
off-shell matrix elements (with $m_c = 1.5~{\rm GeV}$ and
$m_b = 4.75~{\rm GeV}$) convoluted with the CCFM $k_T$-unintegrated gluon 
density of the proton (J2003 set 1\cite{ccfm}). The predictions
are calculated using the Monte Carlo program CASCADE~\cite{cascade}.

\subsection{Monte Carlo Simulation}
Monte Carlo simulations are used to correct for the effects of the
finite detector resolution, acceptance and efficiency.  The Monte
Carlo program RAPGAP\cite{Jung:1993gf} is used to generate low $Q^2$
DIS events for the processes $ep \rightarrow eb\bar{b}X$ and $ep
\rightarrow ec\bar{c}X$. The Monte Carlo program DJANGO\cite{Charchula:1994kf} is used to
 generate light quark ($uds$) events. Both programs combine
 $\cal{O}$($\alpha_s$) matrix elements with higher order QCD effects
 modelled by the emission of parton showers. The heavy flavour event
 samples are generated according to the massive PGF matrix element
 with the mass of the $c$ and $b$ quarks set to $m_c=1.5~{\rm GeV}$
 and $m_b=4.75~{\rm GeV}$, respectively.  In the heavy flavour event
 generation, the DIS cross section is calculated using the parton
 distribution functions (PDFs) from \cite{Martin:1994kn}.  The light
 flavour event samples are generated with the LO PDFs
 from\cite{Gluck:1994uf}.  The partonic system for all generated
 events is fragmented according to the LUND string model implemented
 within the JETSET program\cite{Sjostrand:2000wi}.  The HERACLES
 program\cite{Kwiatkowski:1990es} calculates single photon radiative
 emissions off the lepton line, virtual and electroweak corrections.
 The Monte Carlo program PHOJET\cite{Engel:1995yd} is used to simulate
 the background contribution from photoproduction ($\gamma p
 \rightarrow X$).

The samples of events generated for the $uds$, $c$, and $b$ processes
are passed through a detailed simulation of the detector response
based on the GEANT3 program\cite{Brun:1978fy}, and through the same
reconstruction software as is used for the data. A total of $50$
million $uds$ events, $9$ million $c$ events and $ 1$ million $b$
events were simulated to evaluate the cross sections, corresponding to
luminosities of $90$~${\rm pb}^{-1}$, $160$~${\rm pb}^{-1}$ and
$980$~${\rm pb}^{-1}$, respectively.

\section{H1 Detector}
The analysis is based on a low $Q^2$ sample of $e^+p$ neutral current
scattering events corresponding to an integrated luminosity of $57.4$
${\rm pb}^{-1}$, taken in the years 1999-2000, at an $ep$ centre of
mass energy $\sqrt{s} = 319~{\rm GeV}$, with a proton beam energy of
$920~{\rm GeV}$.

Only a short description of the H1 detector is given here; a full
description may be found in\cite{Abt:1997xv}. A right handed
coordinate system is employed at H1 that has its $z$-axis pointing in
the proton beam, or forward, direction and $x$ ($y$) pointing in
the horizontal (vertical) direction.

Charged particles are measured in the central tracking detector (CTD).
This device consists of two cylindrical drift chambers interspersed
with $z$-chambers to improve the $z$-coordinate reconstruction and
multi--wire proportional chambers mainly used for triggering. The CTD
is situated in a uniform $1.15\,{\rm T}$ magnetic field, enabling
momentum measurement of charged particles over the polar angular
range\footnote{\noindent{ The angular coverage of each detector
component is given for the interaction vertex in its nominal
position.}}  $20^\circ< \theta<160^\circ$.

The CTD tracks are linked to hits in the vertex detector (central
silicon tracker CST)\cite{cst} to provide precise spatial track
reconstruction. The CST consists of two layers of double-sided silicon
strip detectors surrounding the beam pipe, covering an angular range
of $30^\circ< \theta<150^\circ$ for tracks passing through both
layers.   The information on the $z$-coordinate of the CST tracks
is not used in the analysis presented in this paper. For CTD tracks
with CST hits in both layers the transverse distance of closest
approach (DCA) to the nominal vertex in $x$--$y$ can be measured with
a resolution of $33\;\mu\mbox{m} \oplus 90 \;\mu\mbox{m} /p_T
[\mbox{GeV}]$, where the first term represents the intrinsic
resolution (including alignment uncertainty) and the second term is
the contribution from multiple scattering in the beam pipe and the
CST; $p_T$ is the transverse momentum of the track.

The track detectors are surrounded in the forward and central
directions ($4^\circ<\theta<155^\circ$) by a fine grained liquid argon
calorimeter (LAr) and in the backward region
($153^\circ<\theta<178^\circ$) by a lead--scintillating fibre
calorimeter (SPACAL)\cite{Nicholls:1996di} with electromagnetic and
hadronic sections. These calorimeters provide energy and angular
reconstruction for final state particles from the hadronic system. The
SPACAL is used in this analysis to measure and identify the scattered
positron.  A planar drift chamber (BDC~\cite{Adloff:2000qk}),
positioned in front of the SPACAL ($151^\circ<\theta<178^\circ$),
measures the angle of the scattered positron and allows suppression of
photoproduction background, where particles from the hadronic final
state fake a positron signal.

Electromagnetic calorimeters situated downstream in the positron beam
direction allow detection of photons and electrons scattered at very
low $Q^2$. The luminosity is measured from the rate of photons
produced in the Bethe-Heitler process $ep\rightarrow ep\gamma$.

\section{Experimental Method}

\subsection{Event and Track Selection}

The events are selected by requiring a compact electromagnetic cluster
in the SPACAL associated with a track segment in the BDC to define the
scattered positron candidate.  The $z$ position of the interaction
vertex, reconstructed by one or more charged tracks in the tracking
detectors, must be within $\pm 20~{\rm cm}$ of the centre of the
detector to match the acceptance of the CST.  Photoproduction events
are suppressed by requiring $\sum_{i} (E_i - p_{z,i}) >35~{\rm GeV}$.
Here, $E_i$ and $p_{z,i}$ denote the energy and longitudinal momentum
components of a particle and the sum is over all final state particles
including the scattered positron and the hadronic final state
(HFS). The HFS particles are reconstructed using a combination of
tracks and calorimeter deposits in an energy flow algorithm that
avoids double counting. The event kinematics, $Q^2$ and the
inelasticity variable $y$, are reconstructed with the `$e\Sigma$'
method\cite{Bassler:1994uq}, which uses the scattered positron and the
HFS.  The Bjorken scaling variable $x$ is obtained from $x =
Q^2/sy$. In order to have good acceptance in the SPACAL and to ensure that the
HFS has a significant transverse momentum, events are selected in the
range $6.3 < Q^2 < 120 \ {\rm GeV^2}$. The analysis is restricted to
$0.07<y<0.7$ to ensure that the direction of the quark which is struck
by the photon is mostly in the CST angular range.  A further cut of
$y<0.63$ is imposed for events with $Q^2 < 18$~${\rm GeV}^2$ to reduce
photoproduction background.

The triggers used in the analysis require a SPACAL energy deposit in
association with a loose track requirement. Although these triggers
are almost $100\%$ efficient, not all events could be recorded, due to
the large rate for low $Q^2$ events. A fraction of events is rejected at
the first trigger level (L1) and final trigger level (L4).  The Monte
Carlo events are assigned weights to account for the events rejected
at L1 while the data events are assigned weights to account for the
events rejected at L4.  The weights are largest for those events with
an electron at low radius and low energy.
 The overall effect of the trigger is a reduction of the effective
 luminosity by a factor of about $10$ for the lowest $Q^2$ bin and
 $1.4$ for the highest. After applying the event weights and the
 inclusive selection detailed above, the total number of events is
 about $1.5$ million. The background from photoproduction events is
 estimated from the PHOJET Monte Carlo simulation. In most of the $y$
 range this background is negligible and does not exceed $9\%$ in
 any $x$-$Q^2$ bin used in this analysis.

The primary event vertex in $r$--$\phi$ is reconstructed from all
tracks (with or without CST hits) and the position and spread of the
beam interaction region \cite{Aktas:2004az}.  The impact parameter of
a track, which is the transverse distance of closest approach (DCA) of
the track to the primary vertex point, is only determined for those
tracks which are measured in the CTD and have at least two CST hits
linked (referred to as CST tracks). Only CST tracks with a transverse
momentum $>0.5~{\rm GeV}$ are included in the DCA and related
distributions that are used to separate the different quark
flavours. In the kinematic range of this measurement, the fraction of
$c$ ($b$) events that have at least one charged track within the
angular range of the CST, with transverse momentum $>0.5~{\rm GeV}$ and
originating from the decay of a heavy flavoured hadron, is expected to
be $82\%$ ($96\%$), as determined from the Monte Carlo simulation. The
efficiency to obtain a CST track from a CTD track is $76\%$, within
the angular range of the CST.

In order to determine a signed impact parameter ($\delta$) for a
track, the azimuthal angle of the struck quark $\phi_{\rm quark}$ must
be determined for each event. To do this, jets with a minimum $p_T$ of
$2.5 \ {\rm GeV}$, in the angular range $15^\circ < \theta < 155^{\rm
o}$, are reconstructed using the invariant $k_T$ algorithm\cite{KTJET}
in the laboratory frame using all reconstructed HFS particles. The
angle $\phi_{\rm quark}$ is defined as the $\phi$ of the jet with the
highest transverse momentum or, if there is no jet reconstructed in
the event, as $180^\circ-\phi_{\rm elec}$, where $\phi_{\rm elec}$ is
the azimuthal angle of the electron in degrees.  The direction
defined in the transverse plane by $\phi_{\rm quark}$ and the primary
vertex is called the quark axis.
Approximately $81\%$ ($95\%$) of $c$ ($b$) events have $\phi_{\rm quark}$
reconstructed from a jet, as determined from the Monte Carlo simulation.

\begin{figure}[htb]

  \begin{picture}(150,150)
    \put(20,75){\includegraphics[width=0.75\textwidth]{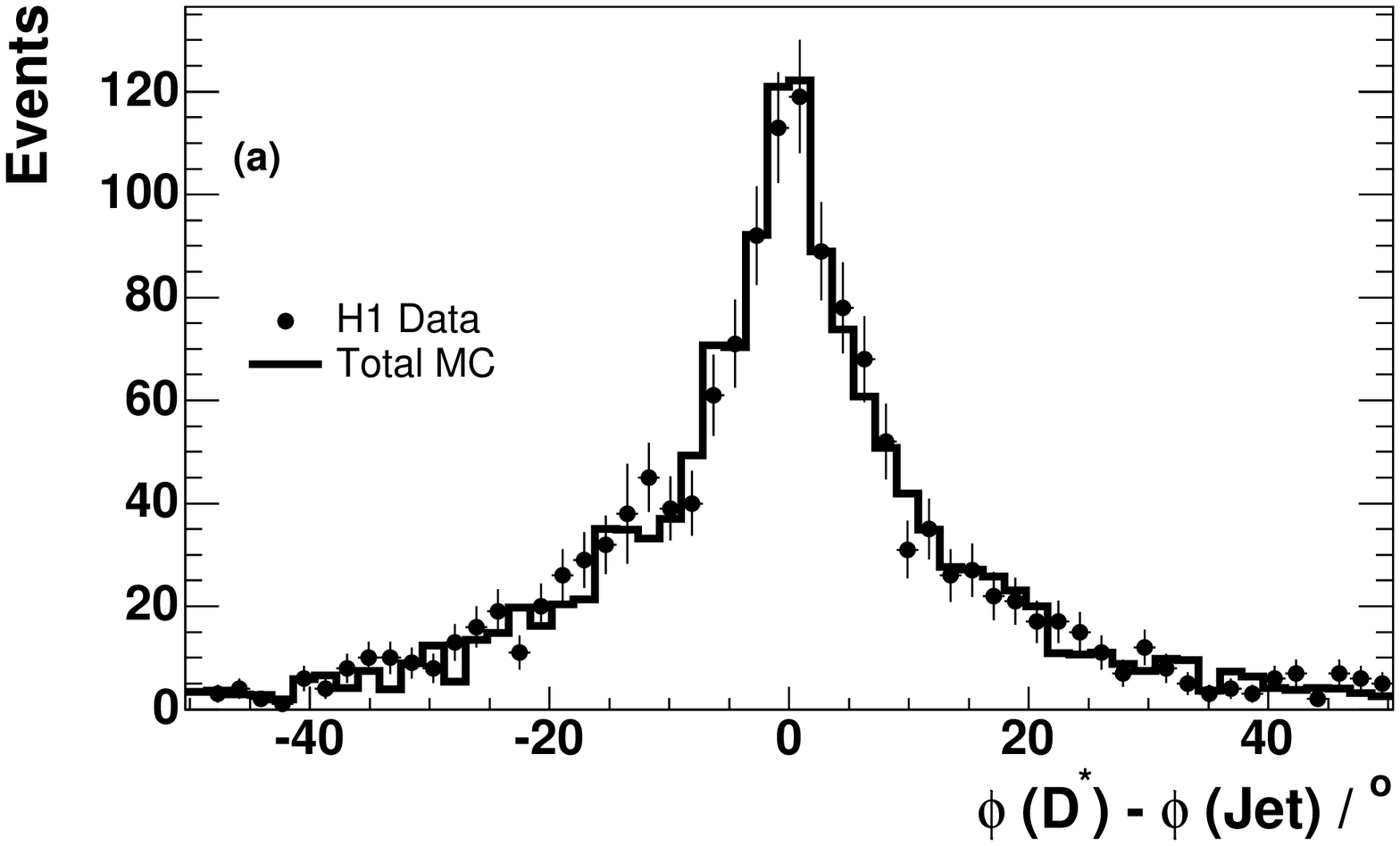}}
    \put(20,0){ \includegraphics[width=0.75\textwidth]{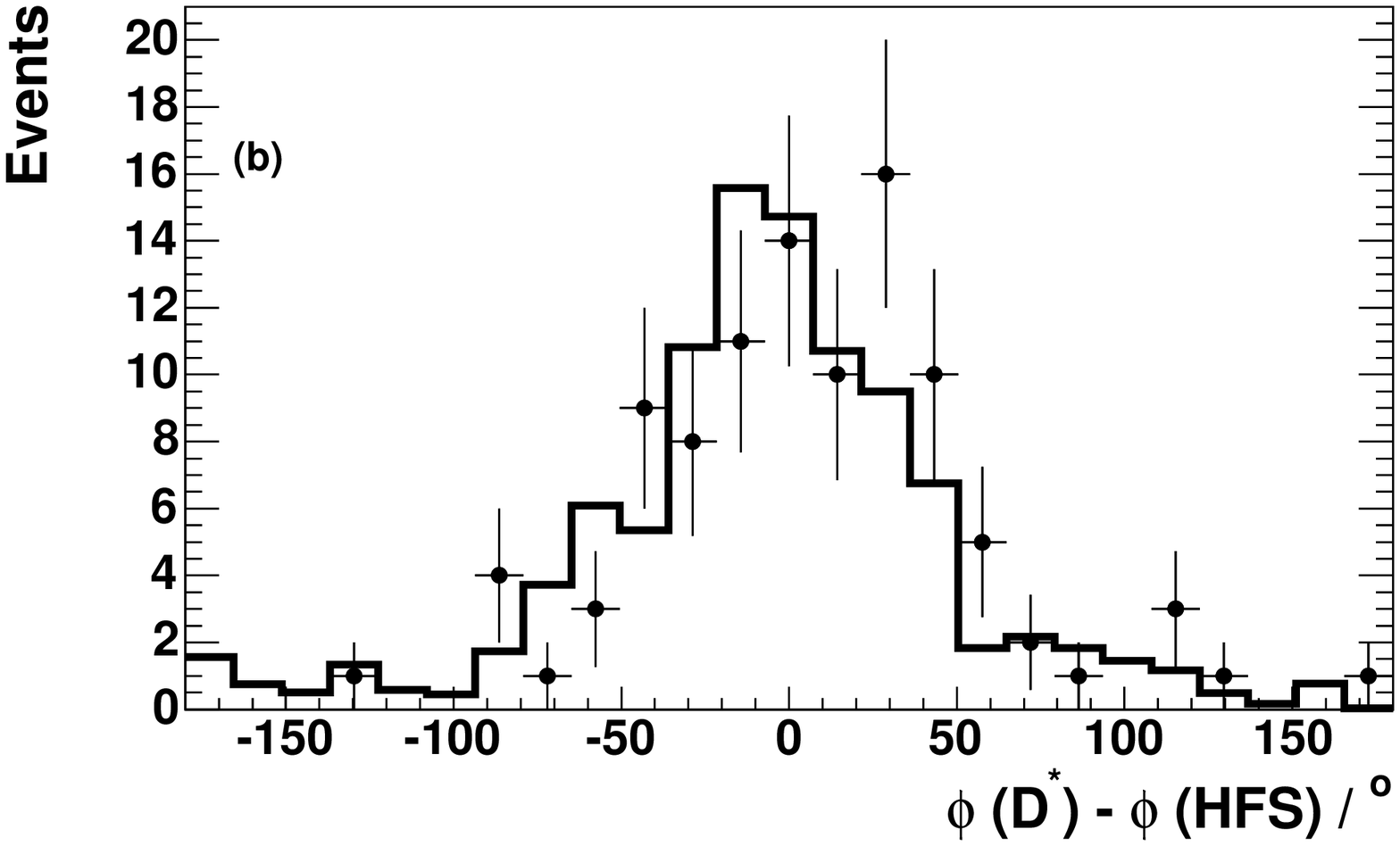}}
 \end{picture}
   \caption{The
  azimuthal difference between the $D^*$ and the quark axis for those
  events where the quark axis is defined (a) by a jet and (b) by
  $180^\circ-\phi_{\rm elec}$. Included in the figure is the
  expectation from the Monte Carlo simulation normalized to the
  number of data events.}  \label{fig:deltaphi}
\end{figure}

The difference between the reconstructed to the true $\phi_{\rm
quark}$ (defined as the azimuthal angle of the quark with highest
transverse momentum) is estimated from the Monte Carlo simulation to
have a resolution of about $5^\circ$ for events with a reconstructed
jet and $35^\circ$ for the rest. The resolution of $\phi_{\rm quark}$
is checked with events containing a reconstructed $D^*$
meson. Figure~\ref{fig:deltaphi} shows the difference between the
reconstructed $D^*$ azimuthal angle and $\phi_{\rm quark}$ for events
with and without a reconstructed jet. Both distributions are well
described by the Monte Carlo simulation.

\begin{figure}[h!]
  \begin{center} \includegraphics[width=0.95\textwidth]{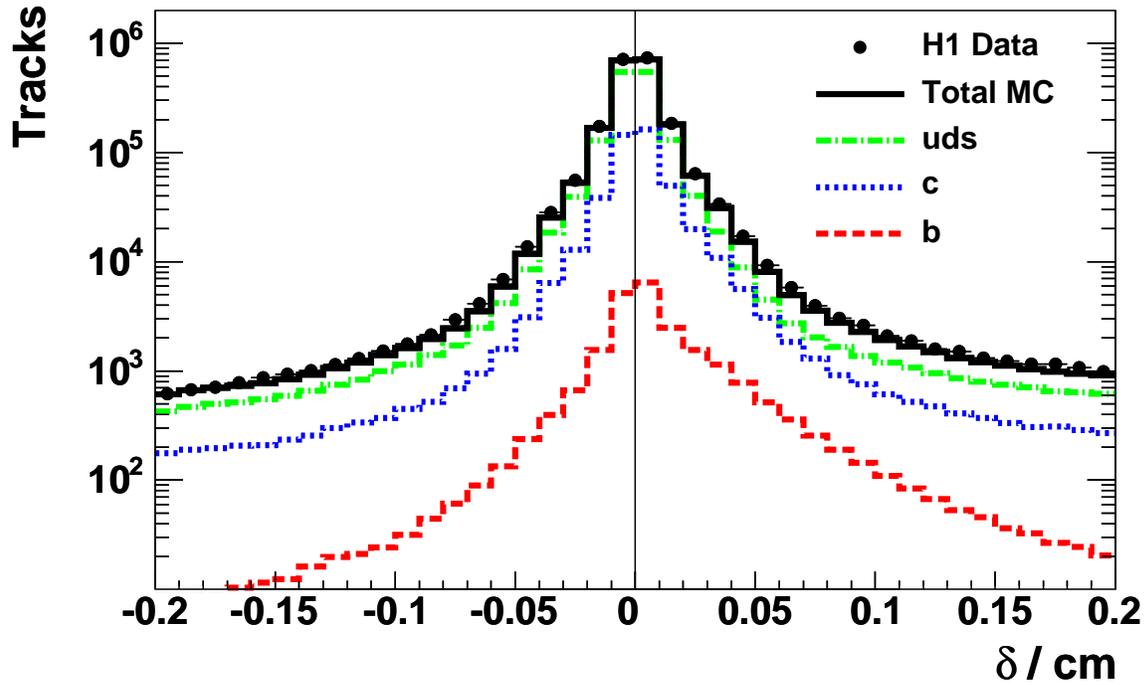}
  \caption{The distribution of the signed impact parameter $\delta$ of
  a track to the primary vertex in the $x$--$y$ plane. Included in the
  figure is the expectation from the DJANGO Monte Carlo simulation for
  light quarks and that from the RAPGAP Monte Carlo simulation for $c$
  and $b$ quarks. The contributions from the various quark flavours
  are shown after applying the scale factors obtained from the fit to
  the subtracted significance distributions of the data (see
  section~\ref{quarkflavourseparation}).}  \label{fig:dca}
  \end{center}
\end{figure}

If the angle between the quark axis and
the line joining the primary vertex to the point of DCA is less than
$90^\circ$, $\delta$ is defined as positive, and is defined as 
negative otherwise. Tracks with
azimuthal angle outside $\pm 90^\circ$ of $\phi_{\rm quark}$ are
rejected. The $\delta$ distribution, shown in figure~\ref{fig:dca}, is
seen to be asymmetric with positive values in excess of negative
values indicating the presence of long lived particles. It is found to
be well described by the Monte Carlo simulation.
Tracks with $|\delta|>0.1~{\rm cm}$ are rejected from the analysis
to suppress light quark events containing long lived strange particles.

\subsection{Quark Flavour Separation}
\label{quarkflavourseparation}

The method used in \cite{Aktas:2004az} to distinguish between the $c$,
$b$ and light quark flavours has been modified in the present analysis
because here the fraction of $b$ quarks is smaller.  The quantities
$S_1$, $S_2$ and $S_3$ are defined as the significance
($\delta/\sigma(\delta)$) of the track with the highest, second
highest and third highest absolute significance, respectively, where
$\sigma(\delta)$ is the error on $\delta$.  Distributions of each of
these quantities are made. The events contributing to the $S_2$
distribution also contribute to the $S_1$ distribution. Similarly,
those contributing to the $S_3$ distribution also contribute to the
$S_2$ and $S_1$ distributions.  Events in which $S_1$ and $S_2$ have
opposite signs are excluded from the $S_2$ distribution. Events in
which $S_1$, $S_2$ and $S_3$ do not all have the same sign are
excluded from the $S_3$ distribution.

\begin{figure}[htb]
  \begin{picture}(150,190)
    \put(20,138){ \includegraphics[width=0.68\textwidth]{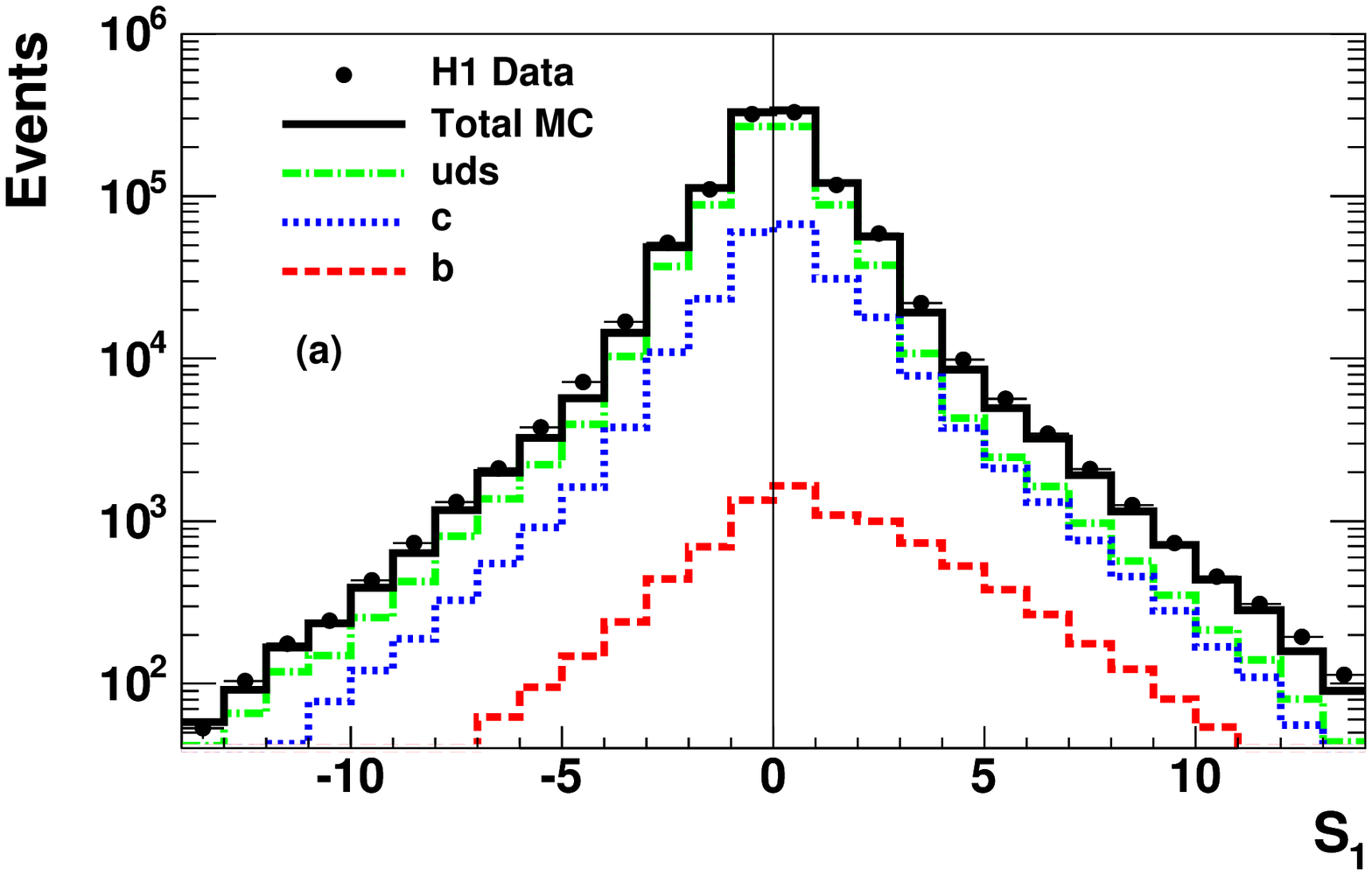}}
    \put(20,69) {  \includegraphics[width=0.68\textwidth]{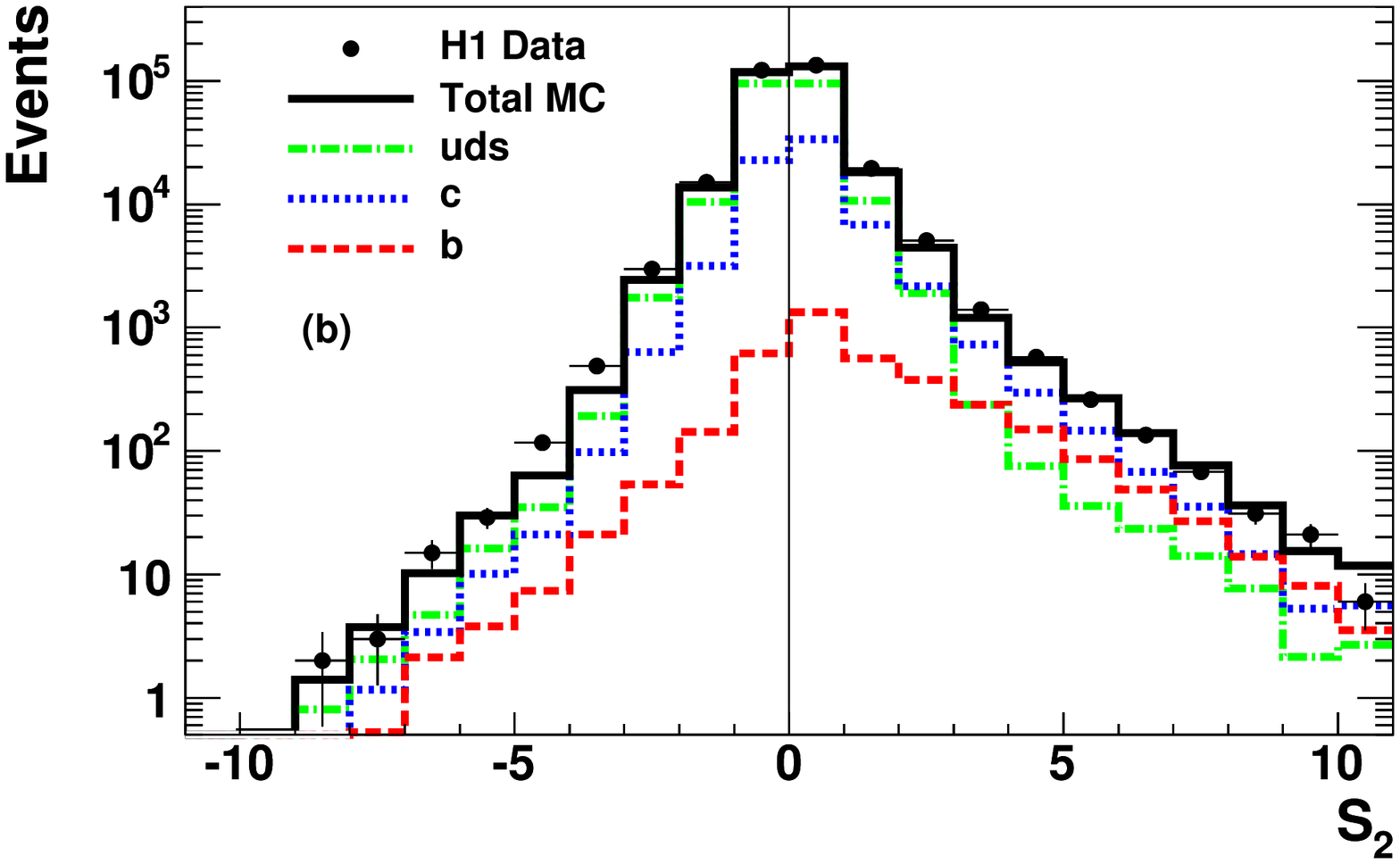}}
    \put(20,0){  \includegraphics[width=0.68\textwidth]{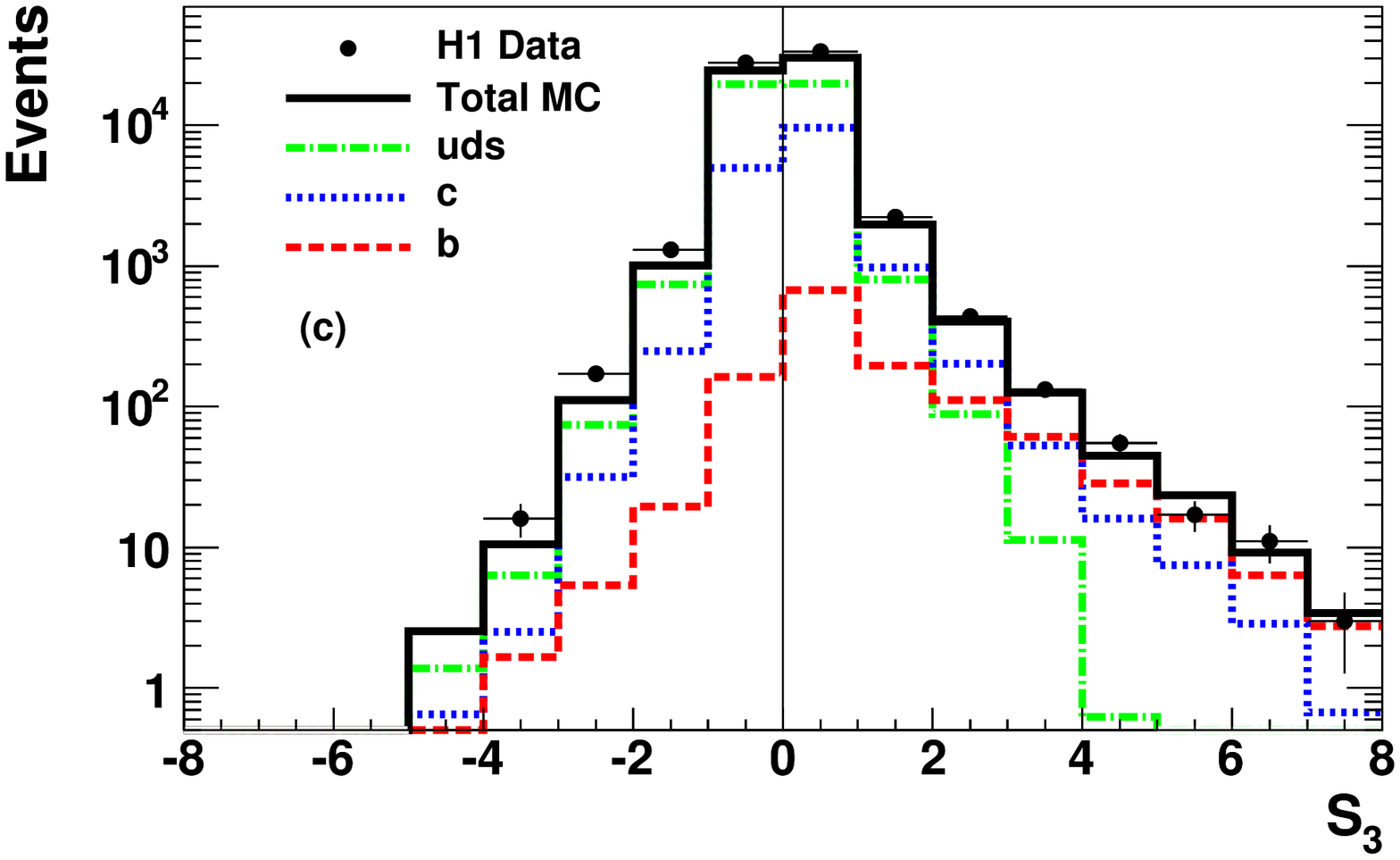}}
 \end{picture}
 \caption{The
  significance $\delta /\sigma(\delta)$ distribution (a) of the
  highest absolute significance track ($S_1$), (b) of the track with
  the second highest absolute significance ($S_2$) and (c) of the
  track with the third highest absolute significance ($S_3$).
  Included in the figure is the expectation from the DJANGO Monte
  Carlo simulation for light quarks and that from the RAPGAP Monte Carlo
  simulation for $c$ and $b$ quarks. The contributions from the various
  quark flavours are shown after applying the scale factors obtained
  from the fit to the subtracted significance distributions of the
  data.}
  \label{fig:s1s2s3}
\end{figure}

\begin{figure}[htb]

  \begin{picture}(150,200)
    \put(15,148){ \includegraphics[width=0.75\textwidth]{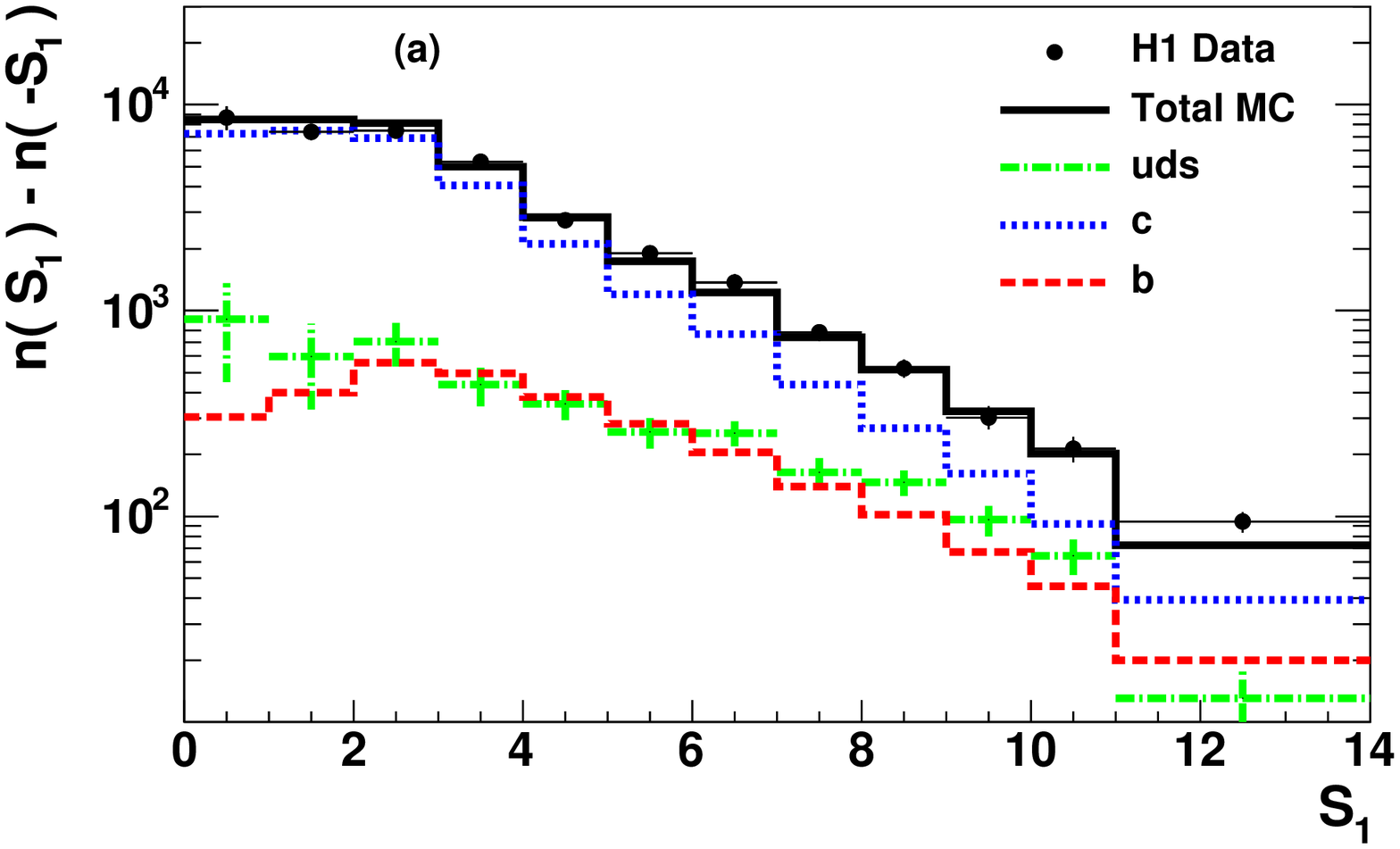}}
    \put(15,74) {  \includegraphics[width=0.75\textwidth]{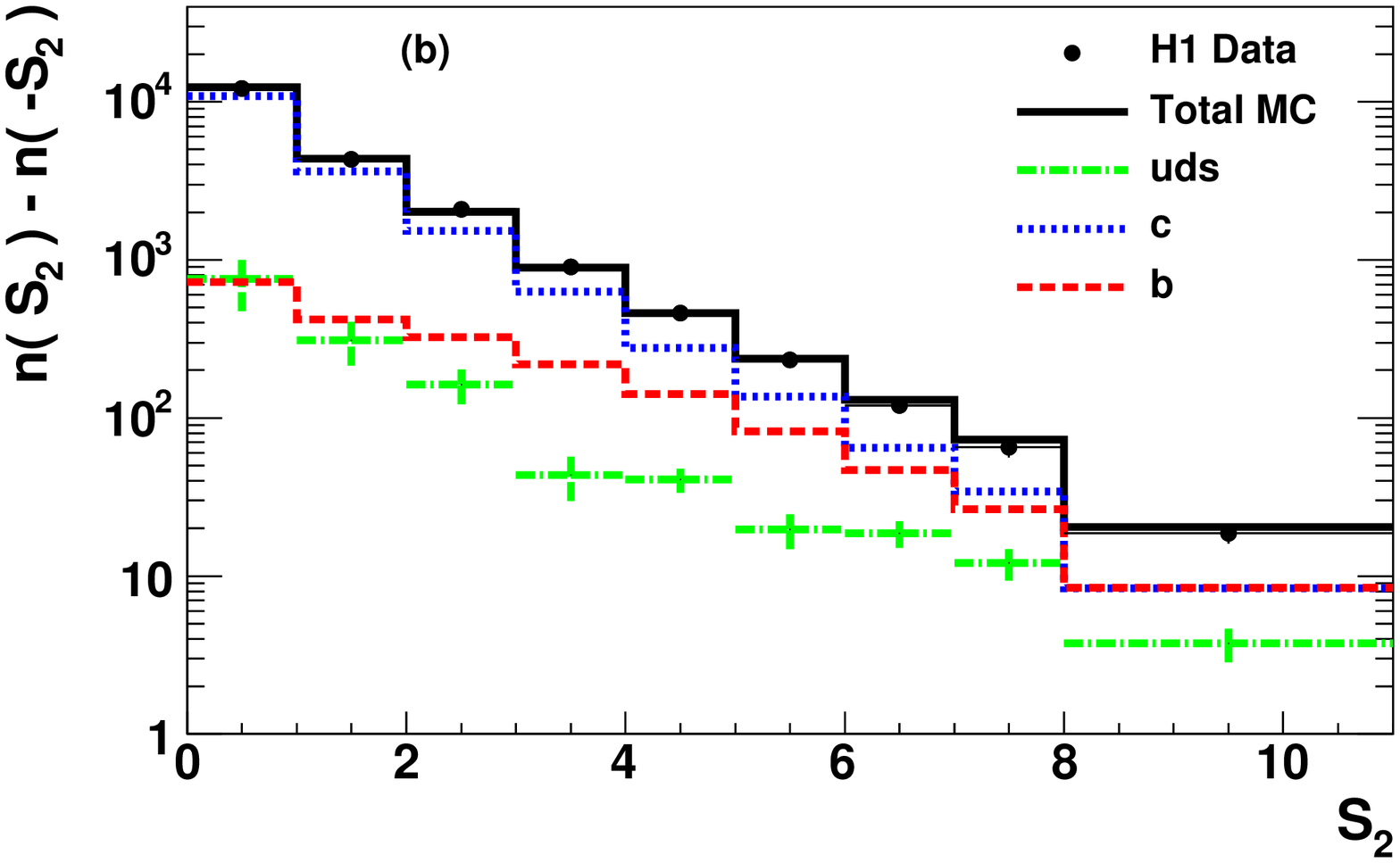}}
    \put(15,0){  \includegraphics[width=0.75\textwidth]{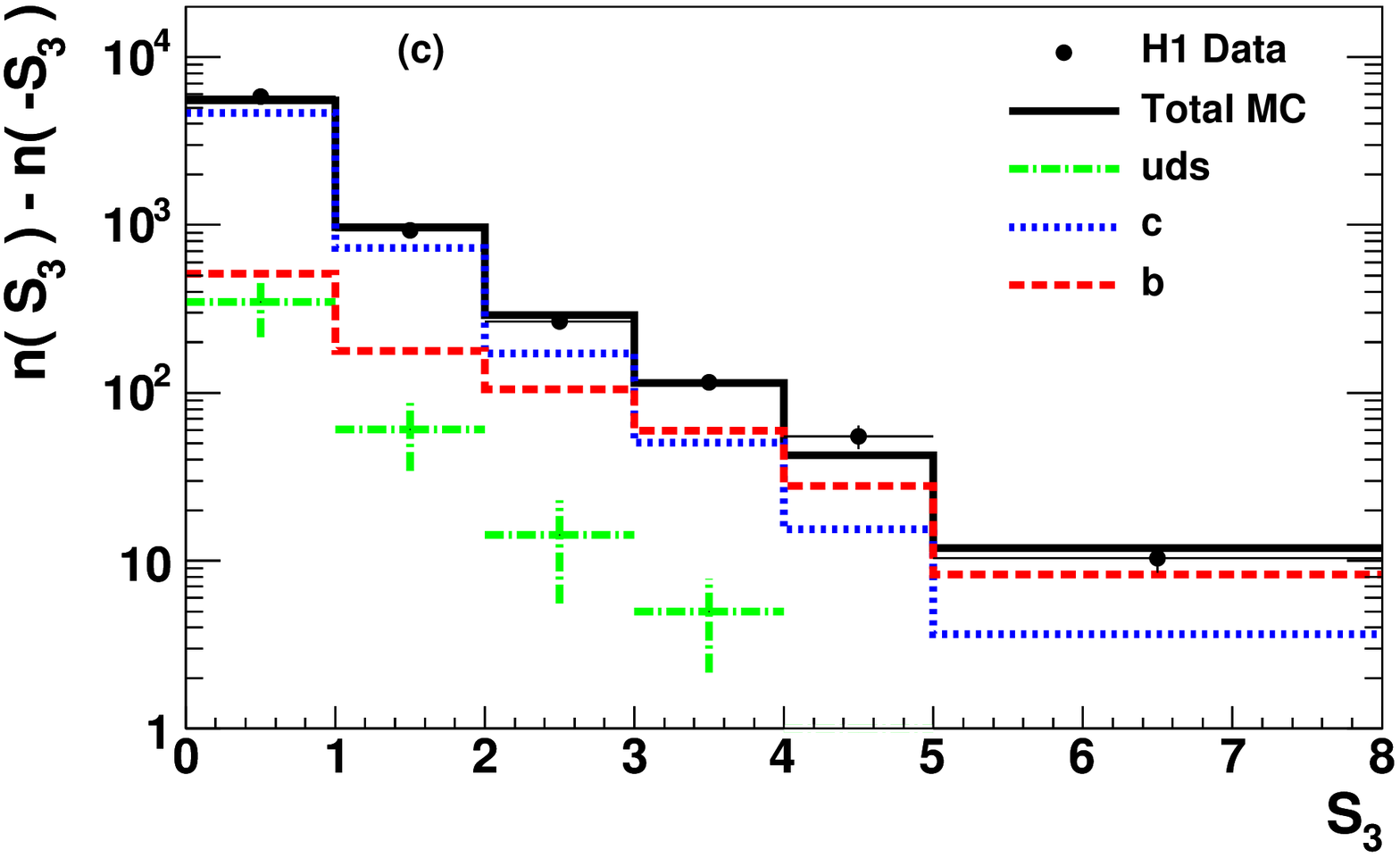}}
 \end{picture}

 \caption{The subtracted
  significance distributions of (a) $S_1$, (b) $S_2$  (c) $S_3$.  Included in the figure is
  the result from the fit to the data of the Monte Carlo distributions
of the various quark flavours.}
  \label{fig:s1s2s3negsub}

\end{figure}

Figure~\ref{fig:s1s2s3} shows the three significance distributions.
The simulation gives a reasonable description of the data. In order to
substantially reduce the uncertainty due to the resolution of $\delta$
and the light quark normalisation, the contents of the negative bins
in the significance distributions are subtracted from the contents of
the corresponding positive bins. The subtracted distributions are
shown in figure~\ref{fig:s1s2s3negsub}. It can be seen that the
resulting distributions are dominated by $c$ quark events, with a $b$
fraction increasing with significance. The light quarks contribute a
small fraction for all values of significance.

The fractions of $c$, $b$ and light quarks of the data are extracted
in each $x$--$Q^2$ interval using a least squares simultaneous fit to
the subtracted $S_1$, $S_2$ and $S_3$ distributions (as in
figure~\ref{fig:s1s2s3negsub}) and the total number of inclusive
events before any CST track selection. The $c$, $b$ and $uds$ Monte
Carlo simulation samples are used as templates.  The Monte Carlo $c$,
$b$ and $uds$ contributions in each $x$--$Q^2$ interval are scaled by
factors $P_c$, $P_b$ and $P_l$, respectively, to give the best fit to
the observed subtracted $S_1$, $S_2$, $S_3$ and total
distributions. Only the statistical errors of the data and Monte Carlo
simulation are considered in the fit.  The fit to the subtracted
significance distributions mainly constrains $P_c$ and $P_b$, whereas
the overall normalisation constrains $P_l$.

The results of the fit to the complete data sample are shown in
figure~\ref{fig:s1s2s3negsub}. The fit gives a good description of all
the significance distributions, with a $\chi^2/ n.d.f$ of $18.0/25$.
Values of $P_c=1.28 \pm 0.04$, $P_b=1.55 \pm 0.16$ and $P_l=0.95 \pm
0.01$ are obtained.  The $c$ and $b$ scale factors are found to be
anti-correlated with an overall correlation coefficient of -0.70.
Acceptable $\chi^2$ values are also found for the
fits to the samples in the separate $x$--$Q^2$ intervals.  Since the
same event may enter the $S_1$, $S_2$ and $S_3$ distributions, it was
checked using a high statistics Monte Carlo simulation that
this has negligible effect on the results of the fits with the present
data statistics.

The results of the fit in each $x$--$Q^2$ interval are converted to a
measurement of the `reduced $c$ cross section'
defined from
the differential cross section as
\begin{equation}
\tilde{\sigma}^{c\bar{c}} (x, Q^2) = \frac{{\rm d}^2\sigma^{c\bar{c}} }{{\rm d} x\,{\rm d} Q^2}  \frac {x Q^4 } {2 \pi \alpha^2 (1+ (1-y)^2)},
\end{equation}

\noindent using:
\begin{equation}
\tilde{\sigma}^{c\bar{c}} (x, Q^2) = 
\tilde{\sigma} (x, Q^2) \frac{P_c N^{\rm MC gen}_c}{P_c N^{\rm MC gen}_c+P_b N^{\rm MC gen}_b+P_l N^{\rm MC gen}_l}  
\delta_{\rm BCC},
\end{equation}
where $\tilde{\sigma} (x, Q^2)$ is the measured inclusive reduced
cross section from H1\cite{Adloff:2000qk} and $N^{\rm MC gen}_c$,
$N^{\rm MC gen}_b$ and $N^{\rm MC gen}_l$ are the number of $c$, $b$ and
light quark events generated from the Monte Carlo in each bin. A bin
centre correction $\delta_{\rm BCC}$ is applied using a NLO QCD
expectation for $\tilde{\sigma}^{c\bar{c}}$ to convert the bin
averaged measurement into a measurement at a given $x$--$Q^2$ point.
The NLO QCD expectation is calculated from the results of a fit
similar to that performed in~\cite{Adloff:1999ah} but using the FFNS
scheme to generate heavy flavours. A small correction ($\le 2.6\%$) 
for the beam energy difference is
applied, using the NLO QCD expectation, to the measurement of
$\tilde{\sigma} (x, Q^2)$ which was performed at a lower centre of
mass energy of $301~{\rm GeV}$ than the data presented here.  The
cross section is defined so as to include a correction for pure QED
radiative effects. Events that contain $c$ hadrons via the decay of
$b$ hadrons are not included in the definition of the $c$ cross
section. The differential $b$ cross section is evaluated in the same
manner.

\subsection{Systematic Errors}
\label{systematics}
\label{sec:systematics}
The systematic uncertainties on the measured cross sections are
estimated by applying the following variations to the Monte Carlo
simulation:
\begin{itemize}
\item An uncertainty in the $\delta$ resolution of the tracks is estimated by varying the resolution by an amount that encompasses
the differences between the data and simulation (figures
\ref{fig:dca}, \ref{fig:s1s2s3}).  This was achieved by applying
an additional Gaussian smearing in the Monte Carlo of
$200$~$\mu{\rm m}$ to $5\%$ of randomly selected tracks and
$25$~$\mu{\rm m}$ to the rest.

\item A  track efficiency uncertainty of $2\%$ due to the CTD and
of $1\%$ due to the CST.

\item The uncertainties on the various $D$ and $B$ meson lifetimes,
  decay branching fractions and mean charge multiplicities are
  estimated by varying the input values of the Monte Carlo simulation
  by the errors on the world average measurements.  For the branching
  fractions of $b$ quarks to hadrons and the lifetimes of the $D$ and
  $B$ mesons the central values and errors on the world averages are
  taken from\cite{Hagiwara:fs}. For the branching fractions of $c$
  quarks to hadrons the values and uncertainties are taken
  from\cite{Gladilin:1999pj}, which are consistent with measurements
  made in DIS at HERA\cite{Aktas:2004ka}. For the mean charged track multiplicities the
  values and uncertainties for $c$ and $b$ quarks are taken from
  MarkIII\cite{Coffman:1991ud} and LEP/SLD\cite{lepjetmulti}
  measurements, respectively.
\item An uncertainty on the fragmentation function of the heavy quarks
  is estimated using the Peterson fragmentation
  function\cite{peterson} with parameters $\epsilon_c = 0.058$ and
  $\epsilon_b = 0.0069$, instead of the LUND fragmentation model.
\item An uncertainty on the QCD model of heavy quark production
 is estimated by using the CASCADE Monte Carlo instead of the RAPGAP Monte Carlo.
\item The uncertainty on the asymmetry of the light quark $\delta$ 
  distribution is estimated by repeating the fits with the subtracted
  light quark significance distributions
  (figure~\ref{fig:s1s2s3negsub}) changed by $\pm50\%$. The light
  quark asymmetry was checked to be within this uncertainty by comparing the
  asymmetry of Monte Carlo events to that of the data, in the
  region $0.1<|\delta|<0.5~{\rm cm}$, where the light quark asymmetry 
  is enhanced.

\item An error on the quark axis is estimated by shifting the 
  quark axis by $2^\circ$($5^\circ$) for events with (without) a
  reconstructed jet. These shifts were estimated by comparing the
  difference between $\phi_{\rm quark}$ and the track azimuthal angle
  in data and Monte Carlo.

\item A $4\%$ uncertainty on the hadronic energy scale.
  
\item Uncertainties on the acceptance and bin centre correction due to
  the input structure functions used are estimated by reweighting the
  input $\tilde{\sigma}^{c\bar{c}}$ distribution by $x^{\pm0.1}$
  and $1 \pm 0.2 \ln [Q^2/(10~{\rm GeV}^2)]$ and
  $\tilde{\sigma}^{b\bar{b}}$ by $x^{\pm 0.3}$ and $1 \pm 0.4 \ln
  [Q^2/(10~{\rm GeV}^2)]$. The range of variation of the input
  structure functions was estimated by comparing to the measured
  values obtained in this analysis.

\item An uncertainty on the photoproduction background is estimated
by assigning $\pm 100\%$ of the expected number of events from the
PHOJET simulation that enter the significance distributions.

\end{itemize}

Other sources of systematic error pertaining to the NC selection were
also considered\cite{Adloff:2000qk}: a $1.5\%$ uncertainty on the
luminosity measurement; an uncertainty on the scattered positron polar
angle of $0.3$~${\rm mrad}$ and energy of $0.3$--$1.0\%$ depending
on the energy; a $0.5\%$ uncertainty on the scattered positron
identification efficiency; a $0.5$--$2\%$ uncertainty on the positron
track-cluster link efficiency; a $\le 1\%$ uncertainty on the trigger
efficiency and a $1\%$ uncertainty on the cross section evaluation due
to QED radiative corrections. 

A detailed list of the systematic effect
on each cross section measurement is given in table~\ref{tab:sig}.
The systematic error is larger for the $b$ measurement than it is for the
$c$ because the $b$ fraction is much smaller than the $c$ fraction.
The errors which contribute most to the uncorrelated systematic error
in table~\ref{tab:sig} are, at low $Q^2$ and high $y$, 
the uncertainty on the photoproduction background and, elsewhere, 
the uncertainty on the acceptance and bin centre correction due
to the input structure function.

\section{Results}
\label{results}
\begin{figure}[htb]
  \begin{center}
  \includegraphics[width=0.89\textwidth]{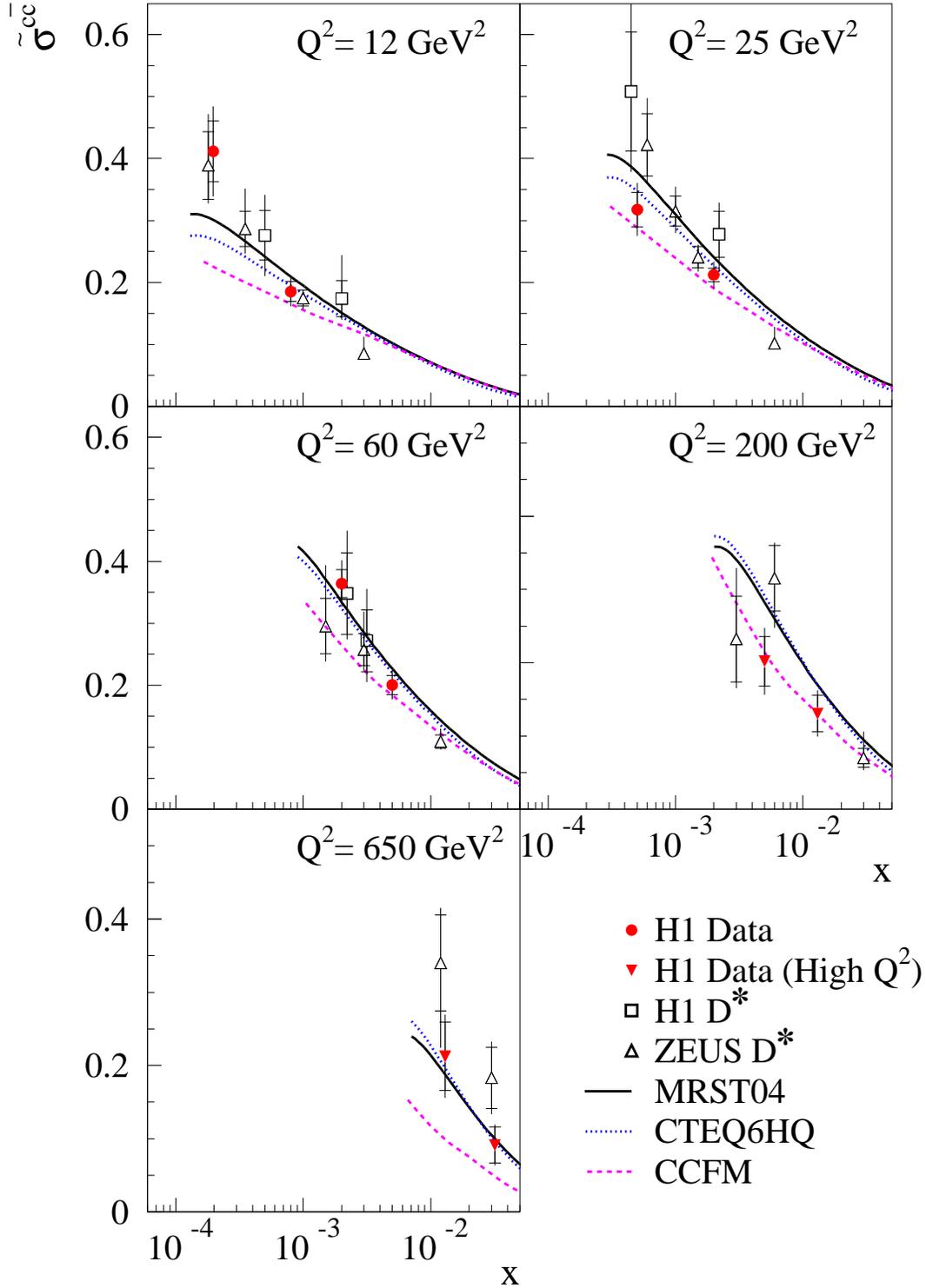} \caption{The
  measured reduced cross section $\tilde{\sigma}^{c\bar{c}}$ shown as
  a function of $x$ for 5 different $Q^2$ values.  The inner error
  bars show the statistical error, the outer error bars represent the
  statistical and systematic errors added in quadrature.  The
  measurements of $\tilde{\sigma}^{c\bar{c}}$ from H1 at high values
  of $Q^2$\cite{Aktas:2004az}, the measurements obtained from $D^*$
  mesons from H1 and ZEUS\cite{H1Dstar,ZEUSDstar} and  predictions
  of QCD are also shown.}  \label{fig:f2cc} \end{center}
\end{figure}

The measurements of $\tilde{\sigma}^{c\bar{c}}$ are listed in
table~\ref{tab:sig} and shown in figure~\ref{fig:f2cc} as a function
of $x$ for fixed values of $Q^2$. The H1 data for
$\tilde{\sigma}^{c\bar{c}}$ are compared with the results extracted
from $D^*$ meson measurements by H1~\cite{H1Dstar} and
ZEUS~\cite{ZEUSDstar} obtained using a NLO program~\cite{Harris:1997zq} 
based on DGLAP evolution to extrapolate the
measurements outside the visible $D^*$ range. 
The measurements for $\tilde{\sigma}^{c\bar{c}}$ from  the present
analysis and the $D^*$ extraction methods are in good agreement.

The $\tilde{\sigma}^{c\bar{c}}$ data are compared with two VFNS predictions from NLO QCD
(see section~\ref{sec:theory}) from MRST\cite{Martin:2004dh} and
CTEQ\cite{cteqvfns},
and with predictions based on CCFM\cite{ccfm2} parton evolution.
The predictions provide
a reasonable description of the present data.

\begin{figure}[htb]
  \begin{center}
  \includegraphics[width=0.89\textwidth]{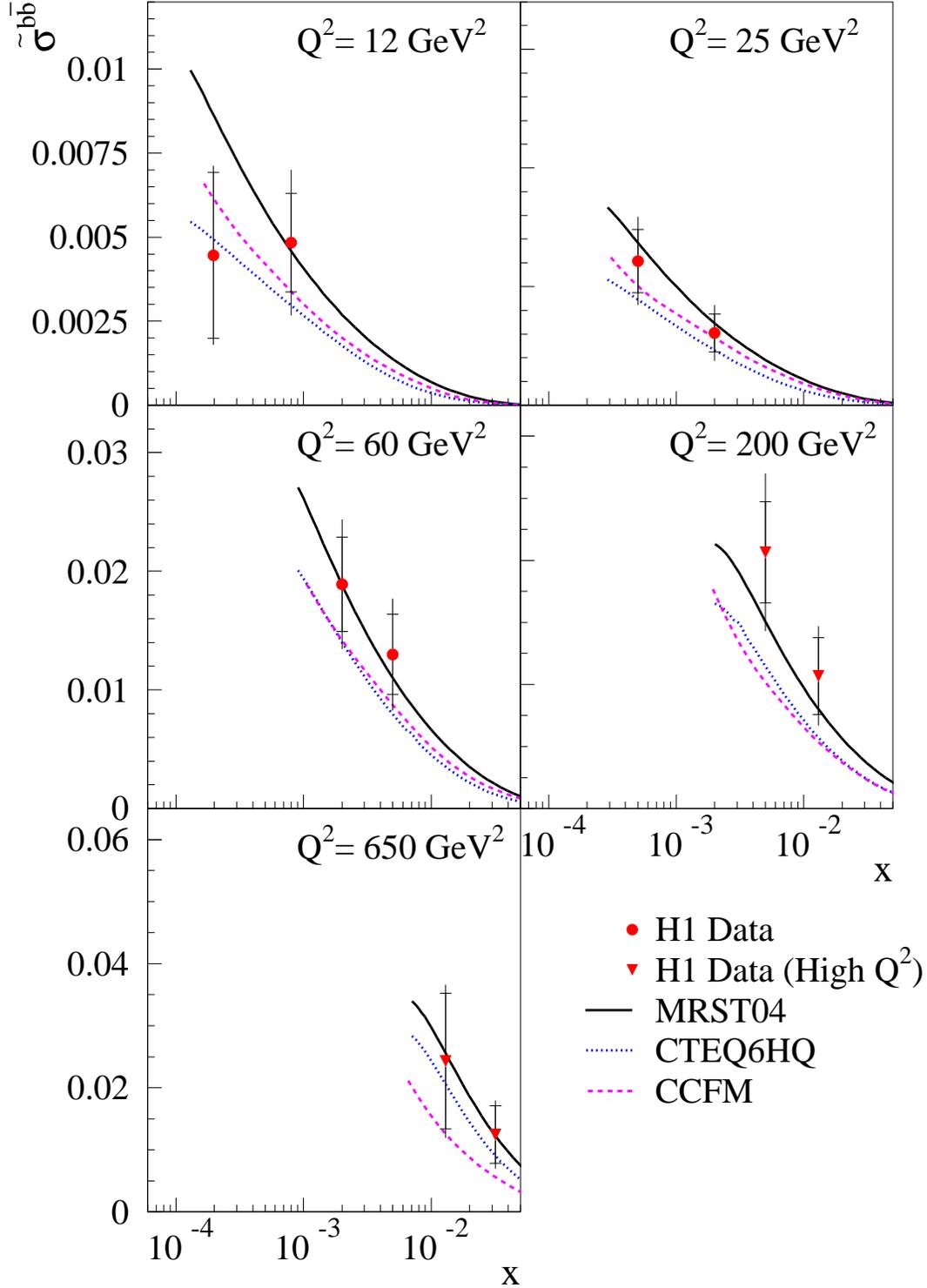} \caption{The
  measured reduced cross section $\tilde{\sigma}^{b\bar{b}}$ shown as
  a function of $x$ for 5 different $Q^2$ values.  The inner error
  bars show the statistical error, the outer error bars represent the
  statistical and systematic errors added in quadrature.  The
  measurements of $\tilde{\sigma}^{b\bar{b}}$ from H1 at high values
  of $Q^2$\cite{Aktas:2004az} and  predictions of QCD are also
  shown.}  \label{fig:f2bb} \end{center}
\end{figure}

The measurements of $\tilde{\sigma}^{b\bar{b}}$ are also listed in
table~\ref{tab:sig} and are shown in figure~\ref{fig:f2bb} as a
function of $x$ for fixed values of $Q^2$.  This is the first measurement
of $\tilde{\sigma}^{b\bar{b}}$ in this kinematic range.
The $\tilde{\sigma}^{b\bar{b}}$ data are also compared with the two  VFNS NLO QCD
predictions and the CCFM prediction.  The difference between
the two VFNS NLO QCD calculations, which reaches a factor $2$ at 
the lowest $Q^2$ and $x$, arises from the different treatments of 
threshold effects by MRST and CTEQ.  Within the current experimental errors
these differences cannot be resolved.

The structure function $F_2^{c\bar{c}}$ is evaluated from the reduced cross section

\begin{equation}
\tilde{\sigma}^{c\bar{c}} =   F_2^{c\bar{c}}   - \frac{y^2 }{1+ (1-y)^2}  F_L^{c\bar{c}},
\label{eq:sigcc}
\end{equation}
where the longitudinal structure function $F_L^{c\bar{c}}$ is
estimated from the same NLO QCD expectation as used for the bin centre
correction.  The structure function $F_2^{b\bar{b}}$ is evaluated in
the same manner.

\begin{figure}[htb]
  \begin{center} \includegraphics[width=0.85\textwidth]{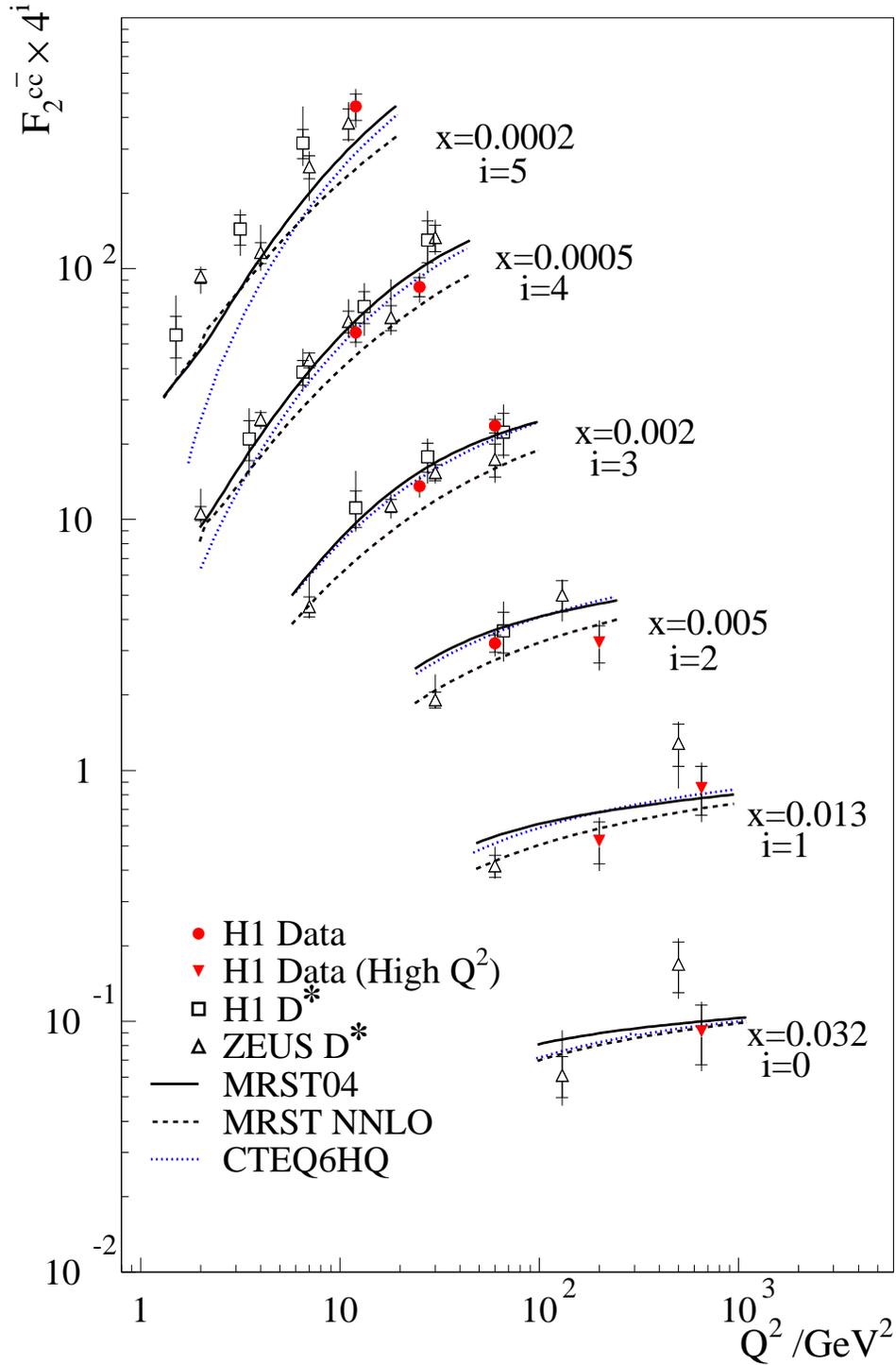}
  \caption{The measured $F_2^{c\bar{c}}$ shown as a function of $Q^2$
  for various $x$ values.  The inner error bars show the statistical
  error, the outer error bars represent the statistical and systematic
  errors added in quadrature. The $F_2^{c\bar{c}}$ from H1 at high
  values of $Q^2$\cite{Aktas:2004az}, the measurements  obtained
  from $D^*$ mesons from H1 and ZEUS\cite{H1Dstar,ZEUSDstar} and 
  predictions of QCD are also shown.}  \label{fig:f2ccq2} \end{center}
\end{figure}

\begin{figure}[htb]
  \begin{center} \includegraphics[width=0.9\textwidth]{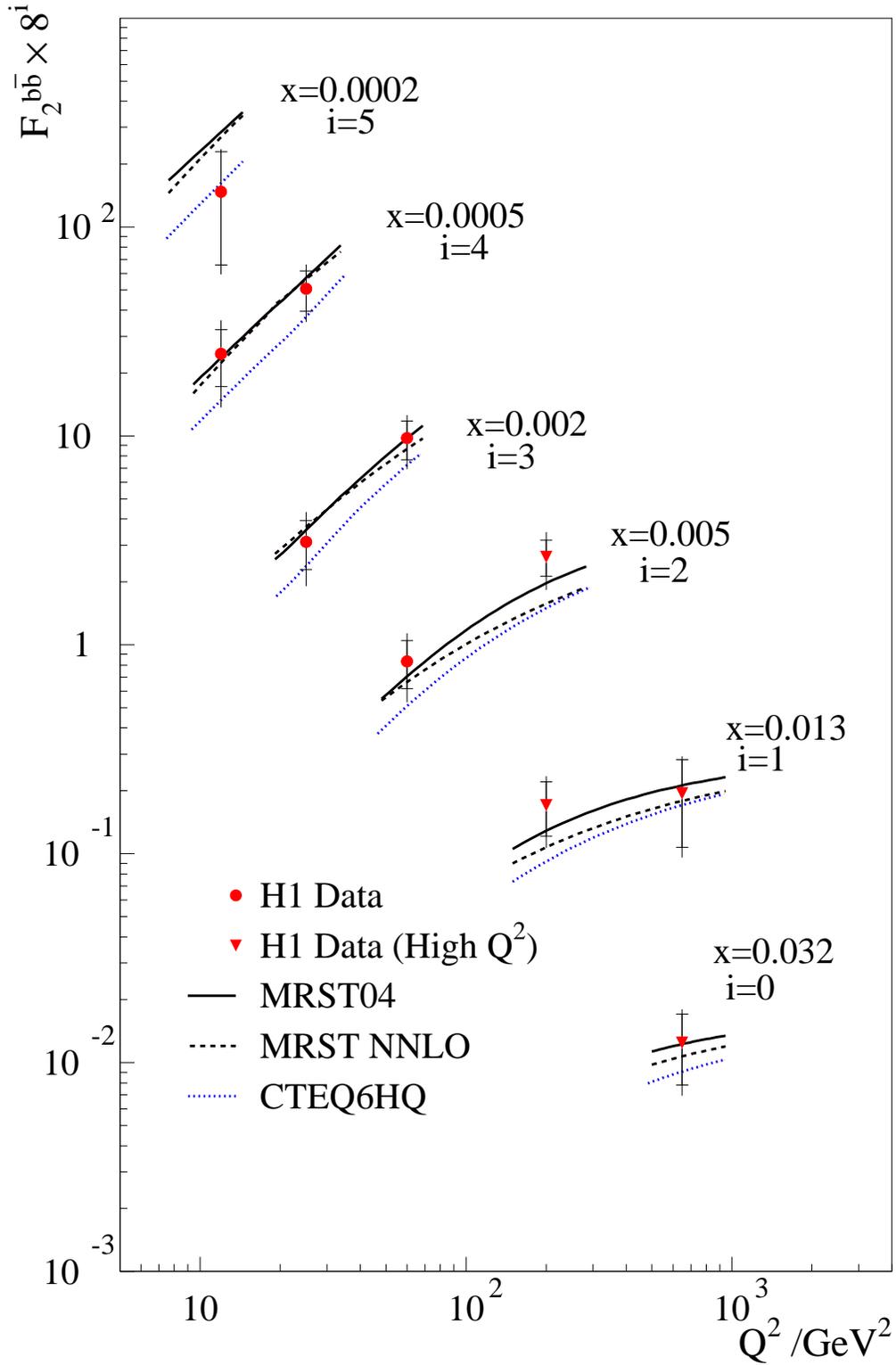}
  \caption{The measured $F_2^{b\bar{b}}$ shown as a function of $Q^2$
  for various $x$ values.  The inner error bars show the statistical
  error, the outer error bars represent the statistical and systematic
  errors added in quadrature. The $F_2^{b\bar{b}}$ from H1 at high
  values of $Q^2$\cite{Aktas:2004az} and  predictions of QCD are
  also shown.}  \label{fig:f2bbq2} \end{center}
\end{figure}

The measurements $F_2^{c\bar{c}}$ and $F_2^{b\bar{b}}$ are shown as a
function of $Q^2$ in figure~\ref{fig:f2ccq2} and
figure~\ref{fig:f2bbq2}.  The measurements of $F_2^{c\bar{c}}$ and
$F_2^{b\bar{b}}$ show positive scaling violations which increase with
decreasing $x$.  The data are compared with the VFNS QCD predictions
from MRST and CTEQ at NLO and a recent calculation at NNLO\cite{NNLO}. The charm data
are more precise than the spread in predictions of the QCD calculations.

\begin{figure}[htb]
  \begin{center} \includegraphics[width=0.9\textwidth]{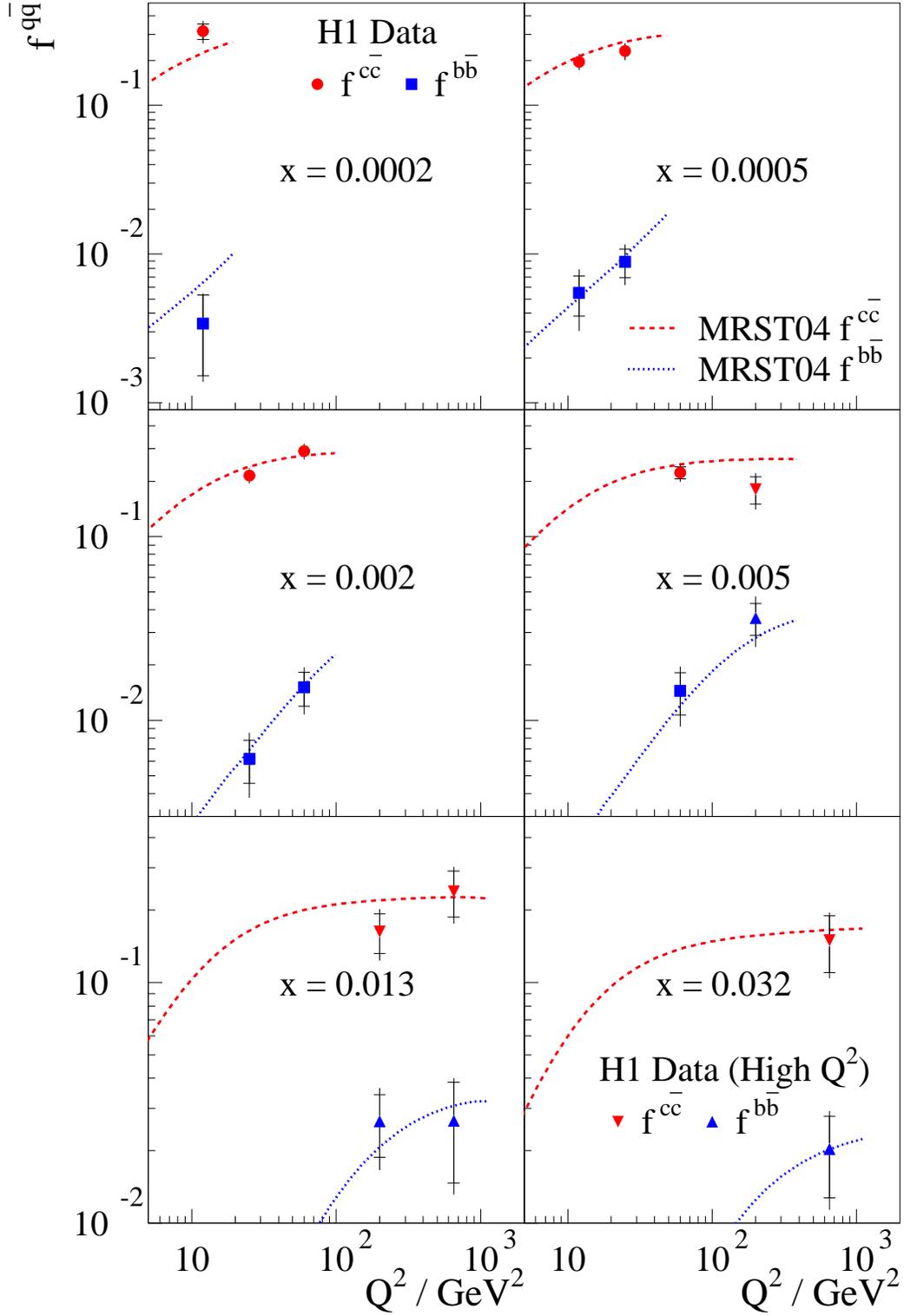} \caption{The
  contributions to the total cross section $f^{c\bar{c}}$ and
  $f^{b\bar{b}}$ shown as a function of $Q^2$ for 6 different $x$
  values. The inner error bars show the statistical error, the outer
  error bars represent the statistical and systematic errors added in
  quadrature.  
  A prediction of NLO QCD is also shown.}
  \label{fig:frac} \end{center}
\end{figure}

The measurements are also presented in table~\ref{tab:frac} and
figure~\ref{fig:frac} in the form of the fractional contribution to
the total $ep$ cross section
\begin{equation}
f^{c\bar{c}} =  \frac{{\rm d}^2 \sigma^{c\bar{c}}} {{\rm d} x\, {\rm d} Q^2}
/
\frac{
{\rm d}^2 \sigma}{ {\rm d} x\, {\rm d} Q^2
}.
\end{equation}
The $b$ fraction $f^{b\bar{b}}$ is defined in the same manner.  
In the present kinematic range the value of 
$f^{c\bar{c}}$ is around $24\%$ on average and 
increases slightly with increasing $Q^2$ and decreasing $x$.
The value of $f^{b\bar{b}}$ increases rapidly with
$Q^2$ from $0.4\%$ at $Q^2 = 12 \ {\rm GeV^2}$ to $1.5\%$ at 
$Q^2 = 60 \ {\rm GeV^2}$. 
The NLO QCD predictions of MRST shown in figure~\ref{fig:frac} are found to
describe the data reasonably well.

\section{Conclusion}

The differential charm and beauty cross sections in Deep Inelastic
Scattering are measured at low $Q^2$ and Bjorken $x$ using the impact
parameters of tracks from decays of long lived $c$ and $b$ hadrons as
reconstructed from the vertex detector.  This is the first measurement
of $F_2^{b\bar{b}}$ in the low $Q^2$ kinematic region.  In this
kinematic range the charm cross section contributes on average $24\%$
of the inclusive $ep$ cross section, and the beauty fraction increases
from $0.4\%$ at $Q^2 = 12 \ {\rm GeV^2}$ to $1.5\%$ at $Q^2 = 60 \
{\rm GeV^2}$.  The cross sections and derived structure functions
$F_2^{c\bar{c}}$ and $F_2^{b\bar{b}}$ are found to be well described
by predictions of perturbative QCD.

\section*{Acknowledgements}

We are grateful to the HERA machine group whose outstanding efforts
have made this experiment possible.  We thank the engineers and
technicians for their work in constructing and maintaining the H1
detector, our funding agencies for financial support, the DESY
technical staff for continual assistance and the DESY directorate for
support and for the hospitality which they extend to the non-DESY
members of the collaboration.  We are grateful to S.~Kretzer,
R.~S.~Thorne and W.~K.~Tung for providing us with their calculations
and for productive discussions.



\begin{sidewaystable}
\small
 \begin{tabular}{|c|c|c|c|c|c|c|c|c|c|c|c|c|c|c|c|c|c|c|} \hline 
  & $Q^2$ & $x$ & $y$ & $\tilde{\sigma}^{q\bar{q}}$ & $C_{cb}$ & $\delta_{\rm stat}$ & $\delta_{\rm sys}$ & $\delta_{\rm tot}$ & $\delta_{\rm unc}$ & $\delta_{\rm res}$ & $\delta_{\rm eff}$ & $\delta_{\rm D mul}$ & $\delta_{\rm B mul}$ & $\delta_{\rm frag}$ & $\delta_{\rm model}$ & $\delta_{uds}$ & $\delta_{\phi}$  & $F_2^{q\bar{q}}$  \bigstrut[t]\\ 
  & (GeV$^2$) & ($\cdot 10^{-3}$) & & & & (\%) & (\%) & (\%) & (\%) & (\%) & (\%) & (\%)  & (\%) & (\%) & (\%) & (\%) & (\%) & \\ \hline
 $c$ &  12 &    0.197 & 0.600 &   0.412 &   -0.62 &    12 &    13 &  18 &    11 &  +3.2 &  -1.4 & -3.1    &  -0.3 &  -0.7 &  -1.9 &  -5.0 &  +2.0 &   0.435 \\ 
 $c$ &  12 &    0.800 & 0.148 &   0.185 &   -0.68 &   8.8 &   9.4 &  13 &   5.6 &  +2.5 &  -1.7 & -3.2   &  -0.2 &  -0.4 &  -2.2 &  -5.2 &  +2.0 &   0.186 \\ 
 $c$ &  25 &    0.500 & 0.492 &   0.318 &   -0.66 &   8.7 &    10 &  13 &   6.8 &  +3.1 &  -1.4 & -3.1   &  -0.3 &  -0.7 &  -1.9 &  -5.0 &  +2.0 &   0.331 \\ 
 $c$ &  25 &    2.000 & 0.123 &   0.212 &   -0.72 &   5.2 &   8.6 &  10 &   4.1 &  +2.6 &  -1.6 & -3.1   &  -0.2 &  -0.5 &  -2.1 &  -5.2 &  +2.0 &   0.212 \\ 
 $c$ &  60 &    2.000 & 0.295 &   0.364 &   -0.74 &   6.2 &   8.3 &  10 &   3.5 &  +3.2 &  -1.4 & -3.1   &  -0.3 &  -0.7 &  -1.9 &  -5.0 &  +2.0 &   0.369 \\ 
 $c$ &  60 &    5.000 & 0.118 &   0.200 &   -0.76 &   7.8 &   8.5 &  12 &   3.8 &  +2.7 &  -1.6 & -3.1   &  -0.2 &  -0.5 &  -2.1 &  -5.1 &  +2.0 &   0.201 \\ 
\hline 
 $b$ &  12 &   0.197 & 0.600 &  0.0045 &   -0.62 &   55 &  22 &  60 &    12 & -13 &  -7.5 &  -2.9 &  +3.0 &  +4.6 &  +8.9 &  -4.8 &  +1.3 &  0.0045 \\ 
 $b$ &  12 &   0.800 & 0.148 &  0.0048 &   -0.68 &   30 &  33 &  45 &    13 & -21 &   -10 &  -5.4 &  +3.1 &  +6.9 & +15 &  -7.7 &  +1.7 &  0.0048 \\ 
 $b$ &  25 &   0.500 & 0.492 &  0.0122 &   -0.66 &   22 &  21 &  31 &   9.1 & -13 &  -7.6 &  -3.0 &  +3.0 &  +4.7 &  +9.1 &  -4.8 &  +1.3 &  0.0123 \\ 
 $b$ &  25 &   2.000 & 0.123 &  0.0061 &   -0.72 &   26 &  28 &  39 &   9.8 & -18 &  -9.4 &  -4.7 &  +3.1 &  +6.3 & +13 &  -6.8 &  +1.6 &  0.0061 \\ 
 $b$ &  60 &   2.000 & 0.295 &  0.0189 &   -0.74 &   21 &  20 &  29 &   6.2 & -13 &  -7.5 &  -2.9 &  +3.0 &  +4.6 &  +8.8 &  -4.7 &  +1.3 &  0.0190 \\ 
 $b$ &  60 &   5.000 & 0.118 &  0.0130 &   -0.76 &   26 &  25 &  36 &   7.4 & -16 &  -8.8 &  -4.1 &  +3.0 &  +5.8 & +12 &  -6.1 &  +1.5 &  0.0130 \\ 
\hline 
\end{tabular} 

\normalsize
    \caption{ The measured reduced NC cross section ($\tilde{\sigma}^{q\bar{q}}$)
    for charm ($c$) and beauty ($b$) quarks, shown with the
    correlation coefficients ($C_{cb}$), the statistical error
    ($\delta_{\rm stat}$), the systematic error
    ($\delta_{\rm sys}$), the total error
    ($\delta_{\rm tot}$) and the uncorrelated
    systematic error ($\delta_{\rm unc}$).  The next
    $8$ columns represent a $+ 1 \sigma$ shift for the correlated
    systematic error contributions from: track impact parameter
    resolution, track efficiency, $D$ multiplicity, $B$ multiplicity,
    fragmentation, QCD model, light quark contribution and quark axis
    $\phi_{\rm quark}$.  The $-1 \sigma$ errors are taken as
    the negative of the upward errors.  The errors are correlated
    between charm and beauty but uncorrelated to inclusive data, apart
    from a normalisation uncertainty of $1.5\%$ which is $100\%$
    correlated. The table also shows the values for $F_2^{c\bar{c}}$
    and $F_2^{b\bar{b}}$ obtained from the measured cross sections
    using the NLO QCD fit to correct for the contributions from
    $F_L^{c\bar{c}}$ and $F_L^{b\bar{b}}$. The quoted relative errors
    apply also to $F_2^{c\bar{c}}$ and $F_2^{b\bar{b}}$.}
\label{tab:sig} 
\end{sidewaystable}

\newpage

\begin{table}
 \begin{tabular}{|c|c|c||c|c|c|c||c|c|c|c|} \hline 
 $x$ & $y$ & $Q^2$ & $f_{c\bar{c}}$ & $\delta^{c\bar{c}}_{\rm stat}$ & $\delta^{c\bar{c}}_{ \rm sys}$ &   $\delta^{c\bar{c}}_{ \rm tot}$ &  $f_{b\bar{b}}$ & $\delta^{b\bar{b}}_{\rm stat}$ & $\delta^{b\bar{b}}_{\rm sys}$ &  $\delta^{b\bar{b}}_{ \rm tot}$  \bigstrut[t] \\ 
 & & (GeV$^2$) & & (\%)  &  (\%)  &  (\%) &  & (\%) &  (\%) &  (\%) \\ \hline
 0.000197  &   0.600   &   12 & 0.316  &      12   &  12   & 17     &  0.0034  &    55   &  22 & 60 \\  
 0.000800  &   0.148   &   12 & 0.188  &     8.6   &   9.1 & 12     &  0.0049  &    30   &  33 & 45    \\  
 0.000500  &   0.492   &   25 & 0.232  &     8.7   &   9.8 & 13     &  0.0089  &    22   &  21 & 30    \\  
 0.002000  &   0.123   &   25 & 0.215  &     5.1   &   8.0 & 10    &  0.0062  &    26   &  28 & 38    \\  
 0.002000  &   0.295   &   60 & 0.291  &     6.1   &   8.0 & 10     &  0.0151  &    21   &  20 & 29    \\  
 0.005000  &   0.118   &   60 & 0.223  &     7.7   &   7.8 & 11     &  0.0144  &    26   &  25 & 36    \\  
 \hline 
 \end{tabular}  

    \caption{ The measured charm ($f^{c\bar{c}}$) and beauty
($f^{b\bar{b}}$) fractional contributions to
the total $ep$ cross section, shown with 
statistical ($\delta^{c\bar{c}}_{\rm stat}$, 
$\delta^{b\bar{b}}_{\rm stat}$), systematic ($\delta^{c\bar{c}}_{\rm sys}$,  
$\delta^{b\bar{b}}_{\rm sys}$) and total ($\delta^{c\bar{c}}_{\rm tot}$, $\delta^{b\bar{b}}_{\rm tot}$) errors.}
\label{tab:frac} 
\end{table}

\newpage

\newpage

\end{document}